\documentclass[12pt]{article}

\catcode`\@=11

\@addtoreset{equation}{section}
\def\theequation{\arabic{equation}}

\def\@normalsize{\@setsize\normalsize{15pt}\xiipt\@xiipt
\abovedisplayskip 14pt plus3pt minus3pt%
\belowdisplayskip \abovedisplayskip
\abovedisplayshortskip  \z@ plus3pt%
\belowdisplayshortskip  7pt plus3.5pt minus0pt}
\def\small{\@setsize\small{13.6pt}\xipt\@xipt
\abovedisplayskip 13pt plus3pt minus3pt%
\belowdisplayskip \abovedisplayskip
\abovedisplayshortskip  \z@ plus3pt%
\belowdisplayshortskip  7pt plus3.5pt minus0pt
\def\@listi{\parsep 4.5pt plus 2pt minus 1pt
            \itemsep \parsep
            \topsep 9pt plus 3pt minus 3pt}}

\def\underline#1{\relax\ifmmode\@@underline#1\else
        $\@@underline{\hbox{#1}}$\relax\fi}
\@twosidetrue
\relax

\catcode`@=12

\catcode`\@=11

\def\section{\@startsection{section}{1}{\z@}{3.5ex plus 1ex minus
   .2ex}{2.3ex plus .2ex}{\large\bf}}

\def\ps@headings{\def\@oddfoot{}\def\@evenfoot{}
\def\@oddhead{\hbox{}\hfill
        \makebox[.5\textwidth]{\raggedright\ignorespaces --\thepage{}--
        \hfill }}
\def\@evenhead{\@oddhead}
\def\subsectionmark##1{\markboth{##1}{}}
}

\ps@headings

\catcode`\@=12

\relax

\def\figcap{\section*{Figure Captions\markboth
        {FIGURECAPTIONS}{FIGURECAPTIONS}}\list
        {Fig. \arabic{enumi}:\hfill}{\settowidth\labelwidth{Fig. 999:}
        \leftmargin\labelwidth
        \advance\leftmargin\labelsep\usecounter{enumi}}}
 \relax
\def\tablecap{\section*{Table Captions\markboth
        {TABLECAPTIONS}{TABLECAPTIONS}}\list
        {Table \arabic{enumi}:\hfill}{\settowidth\labelwidth{Table 999:}
        \leftmargin\labelwidth
        \advance\leftmargin\labelsep\usecounter{enumi}}}
 \relax
\def\reflist{\section*{References\markboth
        {REFLIST}{REFLIST}}\list
        {[\arabic{enumi}]\hfill}{\settowidth\labelwidth{[999]}
        \leftmargin\labelwidth
        \advance\leftmargin\labelsep\usecounter{enumi}}}
 \relax

\catcode`\@=11

\def\marginnote#1{}
\newcount\hour
\newcount\minute
\newtoks\amorpm
\hour=\time\divide\hour by60
\minute=\time{\multiply\hour by60 \global\advance\minute by-
\hour}
\edef\standardtime{{\ifnum\hour<12 \global\amorpm={am}%
    \else\global\amorpm={pm}\advance\hour by-12 \fi
    \ifnum\hour=0 \hour=12 \fi
    \number\hour:\ifnum\minute<100\fi\number\minute\the\amorpm}}
\edef\militarytime{\number\hour:\ifnum\minute<100\fi\number\minute}
\def\draftlabel#1{{\@bsphack\if@filesw {\let\thepage\relax
  \xdef\@gtempa{\write\@auxout{\string
    \newlabel{#1}{{\@currentlabel}{\thepage}}}}}\@gtempa
    \if@nobreak \ifvmode\nobreak\fi\fi\fi\@esphack}
     \gdef\@eqnlabel{#1}}
\def\@eqnlabel{}
\def\@vacuum{}
\def\draftmarginnote#1{\marginpar{\raggedright\scriptsize\tt#1}}
\def\draft{\oddsidemargin -.5truein
        \def\@oddfoot{\sl preliminary draft \hfil
        \rm\thepage\hfil\sl\today\quad\militarytime}
        \let\@evenfoot\@oddfoot \overfullrule 3pt
        \let\label=\draftlabel
        \let\marginnote=\draftmarginnote
   
\def\@eqnnum{(\theequation)\rlap{\kern\marginparsep\tt\@eqnlabel}%
\global\let\@eqnlabel\@vacuum}  }
\def\preprint{\twocolumn\sloppy\flushbottom\parindent 1em
        \leftmargini 2em\leftmarginv .5em\leftmarginvi .5em
        \oddsidemargin -.5in    \evensidemargin -.5in
        \columnsep 15mm \footheight 0pt
        \textwidth 250mmin      \topmargin  -.4in
        \headheight 12pt \topskip .4in
        \textheight 175mm
        \footskip 0pt
        
\def\@oddhead{\thepage\hfil\addtocounter{page}{1}\thepage}
        \let\@evenhead\@oddhead \def\@oddfoot{} \def\@evenfoot{} 
}
\def\titlepage{\@restonecolfalse\if@twocolumn\@restonecoltrue\onecolumn
     \else \newpage \fi \thispagestyle{empty}\c@page\z@
        \def\thefootnote{\fnsymbol{footnote}} }
\def\endtitlepage{\if@restonecol\twocolumn \else  \fi
        \def\thefootnote{\arabic{footnote}}
        \setcounter{footnote}{0}}  
\catcode`@=12
\relax

\def\ps@headings{\def\@oddfoot{}\def\@evenfoot{}
\def\@oddhead{\hbox{}\hfill
        \makebox[.5\textwidth]{\raggedright\ignorespaces --\thepage{}--
        \hfill }}
\def\@evenhead{\@oddhead}
\def\subsectionmark##1{\markboth{##1}{}}
}

\ps@headings

\relax

\def\firstpage#1#2#3#4#5#6{
\begin{document}
\begin{titlepage}
\nopagebreak
\title{\begin{flushright}
        \vspace*{-1.8in}
        {\normalsize CERN--TH/97-272}\\[-8mm]
   {\normalsize CPTH--S563.1097}\\[-8mm]
        {\normalsize hep-th/9710078}\\[4mm]
\end{flushright}
\vfill
{#3}}
\author{\large #4 \\[1.0cm] #5}
\maketitle
\vskip -7mm     
\nopagebreak 
\begin{abstract}
{\noindent #6}
\end{abstract}
\vfill
\begin{flushleft}
\rule{16.1cm}{0.2mm}\\[-3mm]
$^{\dagger}${\small Laboratoire Propre du CNRS UPR A.0014.}\\
CERN--TH/97-272,\ CPTH--S563.1097\\
October 1997
\end{flushleft}
\thispagestyle{empty}
\end{titlepage}}

%
%
\newcommand{\ie}{\hbox{\it i.e.}\ }    
\newcommand{\Rank}{{\rm Rank}\ }
\newcommand{\axion}{\hbox{\Large $a$} }
\newcommand{\Zint}{{\mbox{\sf Z\hspace{-3.2mm} Z}}}
\newcommand{\Real}{{\mbox{I\hspace{-2.2mm} R}}}
\newcommand{\be}{\begin{equation}}
\newcommand{\ee}{\end{equation}}
\newcommand{\ba}{\begin{eqnarray}}
\newcommand{\ea}{\end{eqnarray}}
\renewcommand{\sp}{\; , \; \; }

\date{}
\firstpage{3155}{}
{\large\sc U-duality and D-brane Combinatorics } 
{B. Pioline$^{\,a,b}$ and E. Kiritsis$^{b}$}  
{\normalsize\sl
$^a$Centre de Physique Th{\'e}orique, Ecole Polytechnique,$^\dagger$
{}F-91128 Palaiseau, France\\[-3mm]
\normalsize\sl $^b$Theory Division, CERN, 1211 Geneva 23,
Switzerland
}
{We investigate D-brane instanton contributions to $R^4$ couplings
in any toroidal compactification of type II theories. Starting
from the 11D supergravity one-loop four-graviton amplitude computed
by Green,
Gutperle and Vanhove, we derive 
the non-perturbative 
$O(e^{-1/\lambda})$ corrections to $R^4$ couplings 
by a sequence of T-dualities, and 
interpret them as precise configurations of bound states 
of D-branes wrapping cycles of the compactification torus.
D$p$-branes explicitely appear as fluxes on D$(p+2)$-branes,
and as gauge instantons on D$(p+4)$-branes. 
Specific rules for weighting these contributions are obtained,
which should carry over to more general situations. Furthermore,
it is shown that U-duality in $D\le 6$ relates these D-brane
configurations to $O(e^{-1/\lambda^2})$ instantons for which a geometric
interpretation is still lacking.
}

\eject
Although a non-perturbative definition of superstring theory is
still elusive, the discovery of string dualities has made it clear
that this theory should include various $p$-brane objects
in its BPS spectrum. Under compactification, these objects can
wrap on $r$-cycles of the compactification manifold to yield
$(p-r)$-branes in lower dimensions, or instanton configurations
if $r=p+1$
\cite{bbs}. An exact calculation of physical couplings
should take these instanton effects into account, 
but the rules for weighting
them are still largely unknown. In some cases, the constraints of
duality are strong enough to determine the exact non-perturbative
completion of these couplings
\cite{ov,gg,bfkov,apt,berko,kepa}, thereby opening a window on
the string theory instanton rules. This way, D-instantons
\cite{gg},
D-particles
\cite{gv,ggv} 
and $(p,q)$ strings 
\cite{kp}
contributions to $R^4$ couplings
in toroidal compactifications of type II string have been brought
under control, as well as general D-brane contributions to four 
derivative couplings in $K3$ compactifications of the same
\cite{apt}.
In this letter, we want to take a more systematic approach to this
problem, and derive the contributions of general D$p$-brane instantons
to $R^4$ couplings in toroidally compactified type II theory,
by applying a sequence of perturbative T-dualities on the well understood
D-particle contribution. Imposing S-duality will force us to include
additional contributions, the origin of which remains to be
elucidated. 

The general contribution of D-particles to $R^4$ couplings
in toroidally compactified type IIA theory has been
obtained in Ref.\ \cite{ggv} from 
the one-loop scattering amplitude of four supergravitons 
in 11-dimensional supergravity compactified on a $(N+1)$-torus:
\be
\label{a1}
A_4 = 2\pi {\cal V}_{11}
\int_0^\infty {dt\over t^{5/2}} \hat{\sum_{n^I}} e^{-{\pi\over t} 
n^I g_{IJ} n^J }\ ,
\ee
where $g_{IJ}$ is the volume ${\cal V}_{11}$ 
metric of the torus in eleven-dimensional
Planck units. The sum runs over $(N+1)$-uplets of non-zero integers 
(as denoted by the hat over the sum) dual to the momentum of the 
supergraviton running in the loop. In terms of type IIA variables,
the metric $g_{IJ}$ decomposes as 
\be
\label{a2}
ds_{11}^2 = R_{11}^2 (dx^{11} + {\cal A}_i dx^i)^2 + {1\over R^{11}} 
dx^i g_{ij} dx^j\ ,
\ee
where the eleven-dimensional radius $R_{11}$ is related to the type IIA
coupling
\cite{witten1} through $R_{11}=e^{2\phi/3}=\lambda^{2/3}$, and
$g_{ij}$ denotes the metric in the string frame. The Kaluza--Klein 
connection ${\cal A}$ coincides with the Ramond--Ramond (RR) one-form gauge
potential of type IIA superstring (RR potentials will be denoted 
by curl letters in the following). Going to the string frame and
Poisson-resumming on 
$n^{11} \rightarrow m$ (the details of the procedure are explained
at length in Ref.\ \cite{kp}),
the amplitude (\ref{a1}) can be expanded at weak coupling as
\be
\label{a3}
A_4 = 2\zeta(3){\cal V} e^{-2\phi} + 
2{\cal V} \hat{\sum_{n^i}} {1\over n^i g_{ij} n^j}
+4\pi {\cal V} e^{-\phi} \hat{\sum_{m}} \hat{\sum_{n^i}} 
{ |m| \over \sqrt{n^i g_{ij} n^j} }
K_1 \left( 2\pi e^{-\phi} |m| \sqrt{n^i g_{ij} n^j} \right)
e^{2\pi i m n^i {\cal A}_i}\ .
\ee
In the above expression, ${\cal V}$ is the volume of the $N$-torus in string 
units, $\zeta(3)$ is Apery's transcendental number
and $K_s(z)$ is the Bessel K function, which for a
large argument approximates to
\be
\label{a31}
{ |m| \over \sqrt{n^i g_{ij} n^j} }
K_1 \left( 2\pi e^{-\phi} |m| \sqrt{n^i g_{ij} n^j} \right)
\simeq
{ (\pi|m|)^{1/2} \over (n^i g_{ij} n^j)^{3/2} }
e^{- 2\pi e^{-\phi} |m| \sqrt{n^i g_{ij} n^j} }
\left(1+O(e^{\phi} ) \right) \ .
\ee
The expansion in Eq.\ (\ref{a3}) precisely displays the tree-level 
and one-loop field-theoretical perturbative contributions 
to $R^4$ couplings, together with a sum of non-perturbative
instantons that can be interpreted as D0-branes 
(or D-particles) whose
Euclidean world-line winds on minimal cycles of 
the $N$-dimensional compactification torus. 
This is hardly surprising, given the fact that the type IIA
field theory is the dimensional reduction of 11D supergravity
under which the D0-branes are the Kaluza-Klein modes
\cite{witten1}.
Indeed, the action of the instantons 
\be
\label{a8}
 S_{cl} = e^{-\phi} \sqrt{n^i g_{ij} n^j} + i~n^i {\cal A}_i
\ee
is precisely the Born-Infeld action of a D0-brane wrapped
on a cycle $\sum n^i \gamma_i$ of the $N$-torus. The integer charge
$m$ can be interpreted as the number of D0-brane bound together,
and emerges as the momentum of the supergraviton along
the eleven-dimensional circle. 
Moreover, Eq.\ (\ref{a31}) shows that
each D0-instanton background receives an infinite number of
perturbative corrections, whose coefficients are easily obtained
from the asymptotics of the $K_1$ Bessel function.

Equation\ (\ref{a3}) therefore
gives a precise prescription for including the effect of
D0-brane instantons, at least for the case of the above $R^4$
couplings. Even if one could have guessed the form of the
instanton sum, it is by no means clear how one could have obtained
 the precise summation prescription including the ``zero-mode'' factors
in Eq.\ (\ref{a31}) from first principles, not to mention the perturbative
corrections around the instanton background. 
Note that Eq.\ (\ref{a3})  also leaves room
for interpretation, since we could rewrite it as 
a sum over the winding numbers $m^i=m~n^i$ and thereby obscure
the role of bound states of D0-branes. Note however that this would
introduce a Jacobian $\mu(\{m^i\}) = \sum_{D|m^i} 1$, whereas we would not
expect arithmetic functions to enter instanton calculus when
formulated in terms of the natural objects. This
``naturality'' argument is a useful guideline in understanding
instanton calculus rules.

Here we want to generalize this result and investigate the form
of higher-dimensional D-brane corrections, which we will obtain
by T-duality from the above result.
It will be sufficient for our purposes to use a sequence of
T-dualities on one cycle (say the first direction) 
of the compactification torus.
To do this, it is convenient to decompose the $N$-dimensional torus as a
$U(1)$ fibration:
\be
\label{a4}
ds^2 = R^2 (dx^{1} + A_a dx^a)^2 + dx^a g_{ab} dx^b \sp B_a = B_{1a}\ ,
\ee
where $B_{ij}$ denotes the Neveu-Schwarz (NS) two-form. 
T-duality in the NS sector takes the well-known form:
\be
\label{a5} 
R\leftrightarrow 1/R \sp  A_a \leftrightarrow B_b 
\sp B_{ab} 
\leftrightarrow B_{ab} -A_a B_b + B_a A_b
\sp e^{-2\phi} R = {\rm const.} \ ,
\ee
mapping IIA to IIB. Note that in contrast to usual practice
we do {\it not} canonically
reduce the NS two-form $B_{ab}$ on the first circle,
so that our $B_{ab}$ is not inert under T-duality.
In order to write down the action on the RR gauge potentials,
it is convenient to group them  
into an inhomogeneous differential form of even or odd degree:
\be
\label{a6}
{\cal R}=\sum {\cal R}_\alpha=
\left[
\matrix{ 
{\cal A}_i dx^i + {\cal C}_{ijk} dx^i\wedge dx^j\wedge dx^k + \dots
\sp& {\rm type\,  IIA} \cr
\axion
+ {\cal B}_{ij} dx^i\wedge dx^j+
{\cal D}_{ijkl} dx^i\wedge dx^j \wedge dx^k \wedge dx^l
+ \dots
\sp& {\rm type\, IIB}}
\right.
\ee
The action of T-duality can now be written as:
\be
\label{a7}
{\cal R} \leftrightarrow  1\cdot {\cal R} + 1\wedge {\cal R}
\ee 
where the operators $1\cdot$ and $1\wedge$ are the interior and
exterior products with the first direction, for instance 
\be
\label{a77}
1\cdot {\cal C} = {\cal C}_{1ij} dx^i \wedge dx^j \sp
1\wedge {\cal C} =  {\cal C}_{ijk} dx^1\wedge dx^i \wedge dx^j \wedge dx^k
\ee
In particular, D$p$-brane states charged under the RR $(p+1)$-form
potential are mapped to states charged under both the $p$- and $(p+2)$-
forms of the dual theory, therefore to a superposition of
D$(p-1)$- and D$(p+1)$-branes. Note that Eq.\ (\ref{a77}) holds
only at zeroth order in the NS 2-form. Indeed, the lower
components of the RR fields $\axion$ and ${\cal A}_i$ 
have (at the perturbative
level) Peccei--Quinn symmetries, and should therefore be mapped to 
fields with a Peccei--Quinn symmetry as well. However, a
constant ($SL(2,\Real)$) shift $\axion \rightarrow \axion + c$ 
of the type IIB  RR scalar has to be accompanied by a 
transformation of the RR two-form 
${\cal B} \rightarrow {\cal B} - c B$, so that only 
$\tilde{\cal B} = {\cal B} + \axion B$ has a Peccei--Quinn symmetry.
The correct mapping is therefore ${\cal A}_a \rightarrow \tilde{\cal B}_{1a}$,
and ${\cal B}_{ab} \rightarrow {\cal C}_{1ab}$. A similar correction occurs
in the ${\cal C} \rightarrow \tilde{\cal D}
={\cal D} + B\wedge {\cal B} + \axion B \wedge B$ 
transformation.

Our first aim is to study the action of T-duality on the classical action
of the D0-brane of Eq.\ (\ref{a8}). 
Upon dualizing the first direction, we obtain
\be
\label{a9}
 S_{cl} \to e^{-\phi} \sqrt{ ( n^1 + B_{1a} n^a )^2 + n^a~g_{11} g_{ab}~n^b }
+ i~( n^1 \axion + n^a \tilde{\cal B}_{1a} ) \ .
\ee
As it stands, this result is definitely not invariant under 
$SL(N,\Zint)$ reparametrizations
of the $N$-torus. 
It can however be made invariant by reinterpreting $n^1$ as a
scalar charge $n$, and introducing a two-form integer charge 
$n^{ij}=-n^{ji}$ of which $n^a$ is simply the component $n^{1a}$.
The action (\ref{a9}) then takes the form
\be
\label{a10}
S_{cl} = e^{-\phi} \sqrt{ \left( n + {1\over 2} n^{ij} B_{ij} \right)^2
         + {1\over 2} n^{ij}~g_{ik} g_{jl}~n^{kl} }
+ i~ (n \axion + {1\over 2} n^{ij} \tilde{\cal B}_{ij} )\ ,
\ee
and states corresponding to images of D0-branes under T-duality on
the first circle correspond to $n^{ij}=0$ except for $i=1$ or $j=1$.
This condition can be cast in a more intrinsic form by noting that
states obtained from the D0-brane by T-duality on {\it any} circle are
such that $\Rank n^{ij}=2$. Therefore T-duality a priori only
requires a sum over $n, n^{ij}$ such that $n^{ij}$ has rank two (at most).
Note that this restriction is immaterial for $N\le 3$.

We now show that $S_{cl}$ has a natural interpretation as the
Born--Infeld action of the type IIB D-string wrapping on two-cycles 
of the internal torus. The D-string is described by $N$ embedding 
coordinates $X^i$ together with a $U(1)$ gauge field $A_\alpha$ living on
the two-dimensional world-volume. Supersymmetric mappings of a two-torus
to a $N$-torus are described by a set of $2N$ integers $N^i_{\alpha}$:
\be
\label{a11}
       X^i = N^i_\alpha \sigma^\alpha\ ,
\ee
where $\sigma^\alpha$ are the coordinates on the D-string worldsheet torus.
The gauge field in two dimensions consists only of its zero-mode part,
and its curvature, being the first Chern class of a U(1) bundle,
has to have integral flux:
\be    F_{\alpha\beta} = n~\epsilon_{\alpha\beta}\ .
\ee
We can therefore evaluate the Born--Infeld action on this configuration
and find (hatted quantities are pulled back from target space to
the world-volume):
\be
\label{a12}
       \int e^{-\phi} \sqrt{\det(\hat G + \hat B + F)} 
= e^{-\phi} \sqrt{ \left( n + {1\over 2} n^{ij} B_{ij} \right)^2
         + {1\over 2}n^{ij}~g_{ik} g_{jl}~n^{kl} } \ ,
\ee
in precise agreement with Eq.\ (\ref{a10}), upon identifying
\be
\label{a121}
n^{ij}=\epsilon^{\alpha\beta} N^i_\alpha N^j_\beta \ .
\ee
The integer two-form $n^{ij}$ is independent from the 
parametrization of the D-string world-sheet, and describes the
homology class of the two-cycle inside the $N$-torus. Note in particular
that $\Rank n^{ij}=2$.
Moreover, the phase in Eq.\ (\ref{a10}) is easily
seen to be reproduced by the topological coupling on the D-string
world-sheet \cite{mrd}:
\be
\label{122}
\int e^{\hat B + F}\wedge \hat{\cal R} = 
\int \left( \hat{\cal B} + \axion (\hat B + F) \right)  =
n \axion + {1\over 2} n^{ij} \tilde {\cal B}_{ij}\ .
\ee
This coupling here appears as a simple consequence of
T-duality. Setting $n^{ij}=0$, the action (\ref{a10}) reduces
to $n$ times the action of a D-instanton $S_{cl}=\axion + i~e^{\phi}$.
The integer flux $n$ can therefore be identified with the
D-instanton charge, and the D-string with $n\ne 0$
as a ``bound state''
\footnote{We shall make a rather loose use of the term ``state'',
keeping with the philosophy that instanton effects in 
dimension $D$ can be seen as loops of physical states in 
dimension $D+1$. This seems to fail for the case of the D-instantons.}
of a D-string with $n$ D-instantons.
This is closely related to proposals 
for bound states of D$p$- and D$(p+4)$-brane \cite{mrd,witten2},
but here occurs between D$p$- and D$(p+2)$-branes due to the
existence of nontrivial fluxes on a torus. The integer $m$ again
corresponds to the number of D-strings bound together, and there
is no sign of the non-abelian nature of the interaction \cite{witten2}
in our result. 

The sum over D0-branes winding around the
cycles of the compactification manifold therefore implies by T-duality
a sum over D-strings wrapping the two-cycles of the same:
\be
\label{a13}
A_4^{D1}=
4\pi{\cal V} e^{-\phi} \hat{\sum_{m}} \hat{\sum_{n, n^{ij}}} 
{ |m| \over \sqrt{ \det( \hat G + \hat B + F ) }}
K_1 \left( 2\pi |m| e^{-\phi} \int \sqrt{\det(\hat G + \hat B +F)} \right)
e^{2\pi i m \int \hat{\tilde{\cal B}} + \axion F}
\ee
As in the case of the D0-brane, 
by virtue of the asymptotic expansion of the $K_1$ Bessel function,
this sum exhibits an infinite series of perturbative corrections around
each D-string background. The interpretation of the ``zero-mode'' part
in front of $e^{-2\pi S_{cl}}$ is by no means clear at this point. 
However, it becomes transparent by going to an alternative description,
namely a sum over $(p,q)$ string world-sheet instantons, which was the
object of Ref.\ \cite{kp}. This description is obtained by performing
a Poisson resummation over the D-instanton flux $n$
(this is similar to the transformation from an instanton vacuum
to a $\theta$ vacuum, but for the fact that the D-instanton is
really a flux and not a gauge instanton).
Under this operation, the Bessel function $K_1(z)$ turns into
$K_{1/2}(z)=e^{-z}\sqrt{\pi\over 2z}$, and we find
\be
\label{a14}
A_4^{D1}=4\pi{\cal V} \hat{\sum_l} \sum_{p \wedge q =1}
\sum_{\Rank n^{ij} = 2}
{ e^{-2\pi l |p + q \tau| \sqrt { n^{ij}~g_{ik}g_{jl}~n^{kl} }
    +2\pi i l n^{ij} ( q {\cal B}_{ij} - p B_{ij} ) }
\over
 \sqrt { n^{ij}~g_{ik}g_{jl}~n^{kl} } }
\ee
where $\tau= \axion + i e^{\phi}$ is the type IIB $SL(2,\Zint)$ modulus.
The term with $(p,q)=(1,0)$ corresponds to the 
one-loop world-sheet instantons 
on the fundamental string:
\be
\label{a15}
A_4^{(1,0)}=4\pi{\cal V}\bar{\sum_{N^i_\alpha}}
{
e^{-2\pi \sqrt{ d^{ij} ~g_{ik}g_{jl}~n^{kl}} -2\pi i d^{ij} B_{ij} }
\over
\sqrt{ d^{ij} ~g_{ik}g_{jl}~d^{kl} } \ ,
}
\ee
where $d^{ij}=\epsilon^{\alpha\beta} 
N^i_\alpha N^j_\beta$ and the sum runs over the 
different $SL(2,\Zint)$ orbits of $N^i_\alpha$ such that the $d_{ij}$
are not all zero. Indeed, it can be checked that the number of
orbits corresponding to a given rank-2 set of $d_{ij}$ with greatest common
divisor $D$ is $\sum_{l|D} l$ (this generalizes the result obtained
in Ref.\ \cite{kp} for the particular cases of $N=2,3$); the sum
over the integers $N^i_\alpha$ modulo $SL(2,\Zint)$ can then be traded for a 
sum over $l$ and the rank 2 $n_{ij}=d_{ij}/l$ integer matrix, 
with Jacobian $l$. This indeed reproduces Eq.\ (\ref{a14}), and justifies
the rather mysterious ``zero-modes'' coefficients.
The $(p,q)$ string therefore appears as a {\it coherent superposition} of
$p$ D-strings with an arbitrary number of D-instantons, and not 
as a superposition of $p$ D-string and $q$ D-instantons as one might have 
naively guessed. Moreover, the $(p,q)$ string background appears
to generate {\it no} perturbative corrections, in contrast 
to what occurs around a D-string background in Eq.\ (\ref{a13}).
These perturbative corrections are effectively summed up by
going to the ``$\theta$ vacuum''.

Having obtained the type IIB D-string instanton effects from the 
knowledge of the type IIA D0-brane contribution, we now want to 
investigate higher-brane effects in type IIA by applying one further
T-duality. A generic D-string--D-instanton 
configuration will now be mapped to a superposition of D0-branes that
we started with and D2-branes, which we shall again discover 
by covariantizing the result under the reparametrization group of the 
$N$-torus. Upon T-dualization of the first circle, the action (\ref{a12})
turns into
\ba
\label{a16}
S_{cl}&=&e^{-\phi} \left( R^2 \left[ n + A_a n^{1a} + 
{1\over 2} ( B_{ab}-A_a B_{1b} + A_b B_{1a} ) n^{ab} \right]^2
+ {1\over 2}R^2 n^{ab} g_{ac}g_{bd} n^{cd}  \right.\nonumber
\\
&& \left.
+ (n^{1a} + n^{ab} B_{1b} ) g_{ac} (n^{1c} + n^{cd} B_{1d} ) \right)^{1/2}
+ i ~ \left(n {\cal A}_1 + n^{1a} {\cal A}_a + {1\over 2}n^{ab} {\cal C}_{1ab}
\right)
\ea
Defining $n^{1}=n, n^{i}=n^{1i}$ and introducing
the three-form integer charge $n^{ijk}$ such that $n^{1jk}=n^{ij}$, 
we can rewrite the above action as
\ba
\label{a17}
S_{cl}&=&e^{-\phi} \sqrt{
\left( n^i + {1\over 2} n^{ijk} B_{jk} \right) g_{il}
\left( n^l + {1\over 2} n^{lmn} B_{mn} \right) 
+ {1\over 6} n^{ijk}~g_{il}g_{jm}g_{kn}~n^{lmn} }\nonumber
\\
&&+ i~ \left( n^i {\cal A}_i + {1\over 6} n^{ijk} {\cal C}_{ijk} \right)
\ea
When $n^{ijk}=0$, we recover
the action for the D0-brane we started with. 
$n_{ijk}\ne 0$ on the other hand corresponds to states  charged under the
type IIA RR three-form, therefore to D2-brane states. The transformation
of the integer charges can be conveniently summarized by defining 
an integer inhomogeneous antisymmetric form 
\vspace{.3mm}
\be
\label{a18}
{\cal N} = \sum {\cal N}_\alpha = \left[
\matrix{
n^i dp_i + n^{ijk} dp_i \wedge dp_j \wedge dp_k + \dots \sp & {\rm type\, IIA}
\cr
n + n^{ij} dp_i \wedge dp_j  + n^{ijkl} dp_i 
\wedge dp_j \wedge dp_k \wedge dp_l+\dots
\sp & {\rm type\, IIB}
} \right.
\ee

\noindent
which transforms under T-duality in the same way as ${\cal R}$, namely
$
{\cal N} \leftrightarrow  1\cdot {\cal N} + 1\wedge {\cal N}
$.
We note that the set of charges obtained by T-duality from 
Eq.\ (\ref{a12}) satisfies the conditions $\Rank {\cal N}_3 =3$
(\ie $\ {\cal N}_3 \wedge {\cal N}_3 =0$) and ${\cal N}_1 \wedge {\cal N}_3=0$.
The imaginary coupling to RR gauge potentials can now be written
as ${\cal N}\cdot{\cal R}$ and is now obviously invariant under T-duality.

We now would like, in the same spirit as before, to identify the actual
D2-brane configuration corresponding to Eq.\ (\ref{a17}). The wrappings 
of the
D2-brane world-volume on the compactification $N$-torus are now described
by a set of $3N$ integers $N^i_{\alpha}$ 
(where now $\alpha=1\dots3$), transforming
as $N$ triplets under the 
reparametrization group $SL(3,\Zint)$ of the three-torus. 
Evaluating
the Born--Infeld action for this configuration leads precisely to
the action (\ref{a17}), upon identifying
\be
\label{a19} n^{ijk} = \epsilon^{\alpha\beta\gamma} 
N^i_{\alpha}N^j_{\beta}N^k_{\gamma} 
\sp
n^{i}= {1\over 2} \epsilon^{\alpha\beta\gamma} 
N_\alpha^i F_{\beta\gamma}\ .
\ee
This identification is in perfect agreement with the conditions
$\Rank {\cal N}_3=3, {\cal N}_1 \wedge {\cal N}_3=0$. 
Again, the D0-brane appears as
a non-trivial flux of the $U(1)$ gauge field on the D2-brane
world-volume. As a side remark, we note that the world-sheet coupling
yielding the imaginary part of Eq.\ (\ref{a17}) is
$\int \hat{\cal C} + F \wedge \hat{\cal A}$, and {\it not} 
$\int \hat{\cal C} + (F+\hat B)\wedge \hat{\cal A}$ as 
claimed in Ref.\ \cite{mrd}.
The latter would conflict not only with T-duality 
but also with gauge invariance, since a gauge transformation of the 
NS two-form $B \rightarrow B+d\lambda$ has to be accompanied
with a transformation of
the RR three-form ${\cal C} \rightarrow {\cal C}+{\cal A}\wedge d\lambda$,
as required by the common eleven-dimensional origin of 
$B={\cal C}^{(11)}_{11ij}$ 
and ${\cal C}={\cal C}^{(11)}_{ijk}+{\cal A}_i B_{jk}
+{\cal A}_j B_{ki}+{\cal A}_k B_{ij}$. 

We furthermore obtain the precise summation prescription:
\be
\label{a20}
A_4^{D2}=
4\pi {\cal V} e^{-\phi} \hat{\sum_{m}} \hat{\sum_{n^i, n^{ijk}}} 
{ |m| \over \sqrt{ \det( \hat G + \hat B + F ) }}
K_1 \left( 2\pi |m| e^{-\phi} \int \sqrt{\det(\hat G + \hat B +F)} \right)
e^{2\pi i m \int{\hat{{\cal C}} + \hat{\cal A}\wedge F} }
\ee
At this point, we have to ask whether this result is consistent with
U-duality. In particular, seen as a M-theory coupling, it should be 
invariant under $SL(N+1,\Zint)$
reparametrizations of the $(N+1)$-torus, as was the
case for the D0-brane contribution of Eq.\ (\ref{a1}), obtained
by Poisson resummation on $m\rightarrow n^{11}$ from
the loop amplitude Eq.\ (\ref{a3}). We therefore carry out the
same operation on Eq.\ (\ref{a20}), and obtain:
\be
\label{a201}
A_4^{D2}=2\pi {\cal V}_{11} \int_0^{\infty} {dt\over t^{5/2}}
\hat{\sum} e^{-{\pi\over t} {\cal M}^2}\ ,
\ee
with
\ba
\label{a202}
{\cal M}^2 &=& R_{11}^2 \left( n^{11} + {\cal A}_i n^i + {1\over 6} n^{ijk} 
\tilde{\cal C}_{ijk} \right)^2 + 
\left(n^i + {1\over 2} n^{ijk} B_{jk} \right){ g_{il}\over R_{11} }
\left(n^l + {1\over 2} n^{lmn} B_{mn} \right) \nonumber\\
&&+ {R_{11}^2 \over 6}
n^{ijk}~{g_{il}g_{jm}g_{kn}\over R_{11}^3} ~n^{lmn}  \ .
\ea
By a now familiar line of reasoning, we discover that eleven-dimensional
covariance forces us to introduce the integer four-form
$n^{11ijk} = n^{ijk}$ in terms of which 
\be
\label{a203}
{\cal M}^2 = \left( n^I + {1\over 6} n^{IJKL} {\cal C}^{(11)}_{JKL} \right)
g_{IM} \left( n^M + {1\over 6} n^{MNPQ} {\cal C}^{(11)}_{NPQ} \right)
+ {1\over 24} n^{IJKL} g_{IM}g_{JN}g_{KP}g_{LQ} n^{MNPQ}\ .
\ee
For $N=3$, that is for seven-dimensional type IIA string theory, 
the above
expression is simply a rewriting of Eq.\ (\ref{a202}) which makes
$SL(3+1,\Zint)$ invariance manifest. Quite satisfyingly, it is
also invariant under the {\it full} $SL(5,\Zint)$ U-duality group. Indeed,
it can be checked (for instance by applying a T-duality on the
type IIB $SL(5,\Real)$ symmetric matrix in Eq.\ (5.12) of 
Ref.\ \cite{kp}) that ${\cal M}^2$ in Eq.\ (\ref{a203}) is the norm of
the integer vector $(n^1,n^2,n^3,n^4,n^{1234})$ under the quadratic
form parametrizing the scalar manifold $SL(5,\Real)/SO(5)$\footnote{
up to a factor ${\cal V}_{11}^{-4/5}$, that would appear by translating
Eq.\ (\ref{a201}) to the Einstein frame.}. The $R^4$ coupling
in Eq.\ (\ref{a201}) can therefore be expressed as a weight-3/2
Eisenstein series for $SL(5,\Zint)$, as already found in Ref.\ \cite{kp}
from the type IIB point of view.

For $N\ge 3$ however, we find {\it more} integer charges in Eq.\ (\ref{a203})
than expected from D0- and D2-brane states in Eq.\ (\ref{a202}). For
six-dimensional type IIA string theory ($N=4$), the D-brane charges
$n^{11}, n^i, n^{ijk}$ together with the extra integer $n^{1234}$
fill in a {\bf 10} representation $(n^I,m_I=\epsilon_{IJKLM}n^{JKLM})$ 
of the $SO(5,5,\Zint)$ U-duality group, in such a way that 
${\cal M}^2$ is duality invariant. The $R^4$ amplitude is therefore
again given by a weight-3/2 Eisenstein series for $SO(5,5,\Zint)$,
proving the conjecture in Ref.\ \cite{kp}. The correct interpretation
of states with non-zero $n^{ijkl}$ is still missing at this stage.
It can easily be checked that they give rise to non-perturbative
effects of order $e^{-1/\lambda^2}$, much smaller that the effects
of D-brane instantons. They however appear on the same footing
as D2-branes from the eleven-dimensional point of view, and 
presumably also originate from the M-theory membrane, although
their quantum numbers $n^{IJKL},n^{I}$ would be more suggestive 
of a 3+1 extended object with a three-form on the world-volume.

Leaving these exotic states aside for now, we can pursue our line of
reasoning one step further, and obtain the D3-brane contribution
by T-duality
from the D2-brane result in
Eq.\ (\ref{a17}). Again, one is led to introduce
an integer four-form $n^{1jkl}=n^{jkl}$, and finds
\ba
\label{a21}
S_{cl}&=&e^{-\phi} \left[
\left(n + {1\over 2}n^{ij} B_{ij} +{1\over 8} n^{ijkl} B_{ij} B_{kl} \right)^2
+ {1\over 2} \left(n^{ij} + {1\over 2} n^{ijkl} B_{kl} \right)
g_{im}g_{jn}  \left(n^{mn} + {1\over 2} n^{mnpq} B_{pq} \right) \right.
\nonumber
\\
&&\left.
+ {1\over 24} n^{ijkl} g_{im}g_{jn}g_{kp}g_{lq} n^{mnpq}  \right]^{1/2}
+ i~\left( n \axion + {1\over 2}
 n^{ij} \tilde{\cal B}_{ij} 
+ {1\over 24} n^{ijkl} \tilde{\cal D}_{ijkl}  \right)\ ,
\ea
together with the constraints ${\cal N}_2\wedge{\cal N}_2={\cal N}_0{\cal N}_4$
and ${\cal N}_2\wedge{\cal N}_4=0$.
This turns out to be the effective action of a D3-brane
with the identifications 
\be
\label{a22}
n^{ijkl} = \epsilon^{\alpha\beta\gamma\delta} 
N^i_\alpha N^j_\beta N^k_\gamma N^l_\delta
\sp
n^{ij} = {1\over 2} \epsilon^{\alpha\beta\gamma\delta} 
N^i_\alpha N^j_\beta F_{\gamma\delta}
\sp
n = {1\over 8} \epsilon^{\alpha\beta\gamma\delta} 
F_{\alpha\beta} F_{\gamma\delta}\ ,
\ee   
The above constraints generalize the condition $\Rank {\cal N}_2 = 0$,
\ie ${\cal N}_2\wedge{\cal N}_2=0$, that we previously found in the absence of
D3-branes.
The D-string therefore appears as a flux on the D3-brane, while the
D-instanton is nothing but a gauge instanton in the world-volume
gauge theory\footnote{Actually, anti-self-dual $U(1)$ connections 
do not exist on a torus with a generic flat metric. What we really
mean here is that the D-instanton number appears as the first
Pontryagin number of the $U(1)$ bundle on the 3-brane world-volume.
I am grateful to P. van Baal for correspondence on this subject.}.
This equivalence between topological invariants of a gauge bundle
and wrapping of D-branes is simply a reflection of 
the isomorphism between the integer cohomology lattice, in which the
characteristic classes take their values, and the integer homology
lattice, which describes the possible wrappings of extended
objects on the manifold.

We could reiterate this reasoning a few times in order to
obtain the contributions of D-4 and higher-branes, but the pattern
is by now clear, and the generalization of Eqs.\ (\ref{a17})
and (\ref{a21})
obvious. Note moreover that for any finite value of $N$, the process
terminates and yields a result invariant under T-duality.
However, it is by no means guaranteed
to be invariant under U-duality. In fact, just as in the type IIA case,
the $SL(2,\Zint)_\tau$ symmetry of type IIB string theory forces
us to introduce extra states, that we can simply obtain by applying
a T-duality on the type IIA states in Eq.\ (\ref{a203}):
\ba
\label{a23}
{\cal M}^2 &=& 
{e^{2\phi}\over {\cal V}} \left(
m + \axion n + {1\over 2} n^{ij} \left({\cal B}_{ij}+\axion B_{ij}\right)
+{1\over 24} \left(n^{ijkl}+ \axion m^{ijkl}\right) {\cal D}_{ijkl}
\right.\nonumber\\
&&\left. \quad\quad \quad
+{1\over 8} \left( n^{ijkl} B_{ij} - m^{ijkl} {\cal B}_{ij} \right) 
\left({\cal B}_{kl} +\axion B_{kl}\right)
\right)^2 \nonumber\\
&&+
{1\over {\cal V}} \left(
n + {1\over 2} n^{ij} B_{ij}
+{1\over 24} m^{ijkl} {\cal D}_{ijkl} 
+{1\over 8} \left( n^{ijkl} B_{ij} - m^{ijkl} {\cal B}_{ij} \right) B_{kl} 
\right)^2 \\
&&+
{1\over {\cal V}} 
\left(n^{ij} + {1\over 2} \left( n^{ijkl}B_{kl} - m^{ijkl} {\cal B}_{kl} 
\right)\right)
g_{im} g_{jn}
\left(n^{mn} + {1\over 2} \left( n^{mnpq}B_{pq} - m^{mnpq} {\cal B}_{pq} 
\right)\right)\nonumber\\
&&+
{1\over {\cal V}} \left( n^{ijkl}+\axion m^{ijkl} \right)
g_{im}g_{jn}g_{kp}g_{lq} \left( n^{mnpq}+\axion m^{mnpq} \right) 
+
{e^{-2\phi}\over {\cal V}} m^{ijkl} g_{im}g_{jn}g_{kp}g_{lq} m^{mnpq}\nonumber
\ea
In the above expression, $n,n^{ij},n^{ijkl}$ are the D-brane charges
in Eq.\ (\ref{a21}), $m$ is the Poisson dual of the integer $m$
in Eq.\ (\ref{a13}), \ie the analog of $n^{11}$, and $m^{ijkl}$
is the image under T-duality of the type IIA exotic four-form
$n^{ijkl}$. Note that in contrast to D-brane charges, the rank
of this form is not changed under T-duality. Under $SL(2,\Zint)_\tau$
duality, the following quantities
transform as a doublet:
\be
\label{a24}
\left(\matrix{{\cal B}_{ij}\cr B_{ij}}\right)\sp
\left(\matrix{m \cr n }\right)\sp
\left(\matrix{n^{ijkl} \cr m^{ijkl}}\right)\sp
\ee
while ${\cal D}_{ijkl}, 
\nu=e^{\phi}/V^{4/N}$ and $\tilde g_{ij}=V^{-2/N}g_{ij}$
are inert. Eq.\ (\ref{a23}) is manifestly invariant under these
combined transformations. Ironically, we find that the description
of the type IIB three-brane, claimed to be a singlet under
$SL(2,\Zint)$, requires the introduction of a doublet of
wrapping charges, while the description of the $(p,q)$ string
only requires a singlet of wrapping charges $n^{ij}$ 
(together with the doublet $m,n$).
Moreover, the action of $SL(2,\Zint)_\tau$ 
duality looks very different from the electric-magnetic 
duality considered in Ref.\ \cite{tse}. It would be very
interesting to understand Eq.\ (\ref{a23}) from brane
dynamics.

In this letter, we have obtained the explicit form of contributions
of D-brane instantons to a particular coupling, namely $R^4$ in
$N=8$ type II superstrings. This coupling is related by supersymmetry
to terms with sixteen fermions, and therefore can only receive
contributions from BPS states breaking one half of the supersymmetry.
Starting from the D0-brane contribution, we found 
by T-duality contributions from D-branes wrapping
supersymmetric cycles of the compactification manifold,
and obtained a definite rule for weighting these objects.
Several points would deserve clarification.{\it Primo},
it is not clear why the Russian doll structure we exhibited, 
in which D$p$-branes are fluxes in D$(p+2)$-branes, 
instantons in D$(p+4)$-branes, etc., does not further break
supersymmetry. {\it Secundo}, it seems that for describing the
bound states of these objects only the $U(1)$ gauge field
on the world-volume is relevant, whereas these bound states
should in principle correspond to normalizable ground
states of a $U(N)$ gauge theory on the world-volume.
{\it Tertio}, the rule for summing higher-brane contributions,
namely summing over the cycles $n^{ijk\dots}$ on which
the $p$-brane wraps, appears to differ from the rule for
the fundamental string, where one sums over the winding
numbers $N^i_\alpha$ modulo the action of the mapping class
group $SL(p+1,\Zint)$ ($p=1$). The two rules differ by a 
Jacobian which involves arithmetic functions of the $n^{ijk\dots}$,
and in this sense, using the fundamental string rule in the
D$p$-brane case would be unnatural, while it is natural
in the $(p,q)$ string case. 
{\it Finally}, one should ask how these rules carry over
to more general couplings. Couplings related by supersymmetry will
definitely exhibit the same instanton contributions, but with
different vertex insertions modifying the prefactor in Eq.\ (\ref{a31}).
Couplings related by supersymmetry to terms with more than 16
fermions will receive contributions from BPS instantons breaking
more that half of the supersymmetries, together with the present ones.
It would be interesting to obtain the prefactors of these instanton effects
from a careful string computation. Eventually, one would also
like to reproduce these results from Matrix Theory\cite{bfss}.

Furthermore, we have shown that U-duality forces
the inclusion of extra states beyond the familiar D-branes.
On the type IIA side, these states appear as D2-branes
boosted along the eleventh dimension, whereas on the type IIB
they are obtained by a $SL(2,\Zint)_\tau$ duality transformation
from the D3-branes. A distinct feature of these states is that
their contributions scale  as $e^{-1/\lambda^2}$, and are much
smaller than the usual D-branes.
It would however be very interesting to elucidate their nature.

\vskip .5cm

\noindent
{\bfseries Acknowledgements : } We are grateful to I. Antoniadis, A. Kehagias,
N. Obers, H. Partouche, J. Russo, T. Taylor, P. van Baal 
for helpful discussions.

\end{document}